\documentclass[12pt]{article}
\usepackage{amssymb}
\usepackage{amsmath,amsfonts,feynmp}
\usepackage{tipa}
\usepackage{stmaryrd}
\usepackage[pdftex, bookmarks=true,colorlinks,linkcolor=red,urlcolor=blue,citecolor=blue]{hyperref}

\makeatletter\@addtoreset{equation}{section}\makeatother

\def\be{\begin{equation}}
\def\ee{\end{equation}}
\def\bea{\begin{eqnarray}}
\def\eea{\end{eqnarray}}

\makeatletter\@addtoreset{equation}{section}\makeatother

\renewcommand{\title}[1]{\vbox{\center\LARGE{#1}}\vspace{5mm}}
\renewcommand{\author}[1]{\vbox{\center#1}\vspace{5mm}}
\newcommand{\address}[1]{\vbox{\center\em#1}}

\begin{document}

\unitlength = .8mm

\begin{titlepage}
\begin{center}
\hfill \\
\hfill \\
\vskip 1cm

\title{ Spatially homogeneous Lifshitz black holes
\\in five dimensional higher derivative gravity }

\vskip 0.5cm
 {Yan Liu\footnote{Email: liu@lorentz.leidenuniv.nl}} 
 
\address{Instituut-Lorentz for Theoretical Physics, Universiteit Leiden,
\\P. O. Box 9506, 2300 RA Leiden, The Netherlands}

\end{center}

\vskip 1.5cm

\abstract{We consider spatially homogeneous Lifshitz black hole solutions in five dimensional higher derivative gravity theories, which can be possible near horizon geometries of 
some systems that are interesting in the framework of gauge/gravity duality.  
We show the solutions belonging to the nine Bianchi classes in the pure $R^2$ gravity.  
We find that these black holes have zero entropy at non-zero temperatures and 
this property is the same as the case of BTZ black holes in new massive gravity 
at the critical point. In the most general quadratic curvature gravity theories, we find new solutions in Bianchi Type I and Type IX cases. }

\vfill

\end{titlepage}


\section{Introduction}

In gauge/gravity duality,  the near horizon geometry in the gravity side is very crucial to understanding the low energy behavior of the dual field theory, especially in the framework of the semi-holographic description \cite{Faulkner:2010tq}.  The simplest geometry which corresponds to a field theory at finite density is the planar Reissner-Nordstr\"om (RN) AdS$_{d+1}$ black hole. The near horizon geometry of the extremal RN AdS black hole is $AdS_2\times R^{d-1}$ and it has been studied in detail in e.g. \cite{Faulkner:2011tm}.  It has a finite entropy at zero temperature which may indicate an instability of the system \cite{Iqbal:2011in}.  Generally when scalar fields or Fermions are present, due to the backreaction of these matter fields the system will have a more stable ground state and the near horizon geometry turns out to be Lifshitz \cite{Kachru:2008yh} at zero temperature in many cases \cite{{Gubser:2009cg},Horowitz:2009ij,Hartnoll:2009ns,{Hartnoll:2010gu},{Cubrovic:2011xm}}.  However, it has been shown in  \cite{Copsey:2010ya,{Horowitz:2011gh}} that the Lifshitz geometry probably is not a good near horizon geometry because of 
 a mild ``singularity".  This motivates people to consider other possible and more interesting near horizon geometries of extremal black holes.

Both $AdS_2\times R^{d-1}$ and Lifshitz geometries have the usual translational symmetries along the spatial directions where the black hole extends, which might not be quite necessary in order to connect these geometries to the real world condensed matter systems. It was shown recently in \cite{Iizuka:2012iv} that one can slightly relax the symmetry requirements on these spatial directions, i.e. from  the usual translational symmetries  to the requirement of homogeneity. In other words, we are interested in the kind of near horizon geometries of extremal black holes which are homogeneous but can be possibly anisotropic in the spatial directions. 

A homogeneous space is a space in which the physics looks the same everywhere.
This indicates that any two points in the space can be connected by the isometry group transformations. We will focus on the $4+1$ dimensional gravity system. 
In this case because the spatial space is three dimensional, we need three linearly independent Killing vectors $\xi_i$ $(i=1, 2, 3)$ as well as a metric which is 
invariant under the isometry group that is generated by these Killing vectors in order to realize a spatially homogeneous space. 
The Killing vectors satisfy the commutation relation $[\xi_i, \xi_j]=c^k_{~ij}\xi_k$ where $c^{k}_{~ij}$ are structure constants. In three dimensions, there are only nine types of such symmetry groups which are described by the Bianchi classification \cite{Ryan:1975jw}. For each case, we can define three one-forms
$\omega^i$ that are invariant under the isometry transformations.   These one forms satisfy 
$d\omega^i=\frac{1}{2}c^{i}_{~jk}\omega^j\wedge \omega^k$.
Thus $ds^2_\Sigma=g_{ij}(t,r)\omega^i\omega^j$ gives a homogeneous space.

In \cite{Iizuka:2012iv}  they considered the metrics of the following form 
\be\label{geo}
ds^2=L^2\big(-e^{2\beta_t r} dt^2+dr^2+\eta_{ij}e^{(\beta_i+\beta_j)r}\omega^i\omega^j\big). 
\ee
where $\beta_t, \eta_{ij}, \beta_i$ are constants and  we can see that 
the spatial space $\{x_i\}~(i=1,2,3)$ that the black hole extends in is homogeneous with three isometries.
Apart from the requirement of homogeneity this system also has a generalized Lifshitz scaling symmetry 
$r\to r+\epsilon$, $t\to t e^{-\beta_t\epsilon}$ while $\omega^i\to \omega^i e^{-\beta_i\epsilon}$ which makes it more interesting to our purpose.  

AdS and Lifshitz cases have the usual  translational invariance (both homogeneous and isotropic) along $x_i$, i.e. $\omega^i=dx_i$ and $c^{i}_{~jk}=0$ and this belongs to the type I class.  When the structure constants $c^{i}_{~jk}$ are nontrivial there are generalized translational symmetries (homogeneity) along the three spatial directions and we call it spatially homogeneous Lifshitz geometry.   
In \cite{Iizuka:2012iv} they mainly realized these geometries (\ref{geo}) in the Einstein gravity coupled to massive gauge fields \cite{Taylor:2008tg,{Kachru:2008yh}}. It is natural to ask if we can find such geometries as well as finite temperature solutions in other kinds of gravity theories.  This is the main motivation of our work. 

Recently there have been many nice results in the developments on higher derivative gravities.  It was shown \cite{Adams:2008zk} that the Lifshitz vacuum geometry is a solution of the quadratic curvature  gravities.  An interesting development is that  the black hole solutions which are asymptotic to Lifshitz geometry were found in \cite{AyonBeato:2009nh} for $2+1$ dimensional new massive gravity (NMG) 
\cite{Bergshoeff:2009hq}. Motivated by this work,  in  \cite{Cai:2009ac} the Lifshitz black hole solution was  also found for $3+1$ dimensional $R^2$ gravity and the generalization to higher dimensional cases can be found in \cite{AyonBeato:2010tm}. Stationary Lifshitz black holes of $R^2$ gravity theory were found in \cite{Sarioglu:2011vz}.
In five dimensions, higher derivative gravity theories are also very interesting as the higher derivative terms can arise in the effective actions derived from string theory. This motivates us to consider if we can find these  five dimensional spatially  homogeneous Lifshitz black hole solutions similar to \cite{Iizuka:2012iv} in the higher derivative gravity theories. 

In the next section, we give the analytic spatially homogeneous Lifshitz black hole solutions in the pure $R^2$ gravity for all the nine Bianchi classes. The thermodynamic properties of these black holes are also discussed. Similar to \cite{Cai:2009ac}, these black holes have zero entropy while non-zero temperature   which is quite similar to BTZ black holes in NMG at the critical points \cite{Liu:2009bk} as well as Schwartzschild black holes in the higher dimensional critical gravity \cite{Lu:2011zk}. It is natural to expect  that our results provide an interesting toy model for quantum gravity in five dimensions. Then in Sec. \ref{s3}, we consider the spatially homogeneous Lifshitz black hole solutions in the most general quadratic curvature gravities.  New solutions are found for Bianchi Type I and IX models.  The entropy of  the Type IX black hole solutions is calculated and it is found to be independent of the temperature which is a feature of AdS$_2$ black hole \cite{Spradlin:1999bn}.  In the last section, we give our conclusion and discussions.

\section{Spatially homogeneous Lifshitz black holes in $R^2$ gravity}

Motivated by \cite{AyonBeato:2009nh},  it was first observed in \cite{Cai:2009ac} that the constant curvature solutions are always solutions of four dimensional $R^2$ gravity at special coupling. Later it was generalized to higher dimensional case \cite{AyonBeato:2010tm}. Since the solution founds in  \cite{Iizuka:2012iv} are all constant curvature solutions, it is interesting to study them in this kind of $R^2$ gravity.  

We start from the action of the general $f(R)$ gravity \cite{Cai:2009ac,{AyonBeato:2010tm}}
\begin{equation}
\mathcal{S} =\frac{1}{2\kappa^2}\int d^{d+1}x \sqrt{-g}  f(R).
\end{equation}
The equation of motion is 
\be\label{e1}
f' R_{\mu\nu}-\frac{1}{2}g_{\mu\nu} f-(\nabla_\mu\nabla_\nu-g_{\mu\nu}\nabla^2)f'=0
\ee
where $f'=\partial_R f.$

For $f(R)=-\frac{1}{8\lambda}(R-4\lambda)^2h(R)$ with $h(R)$ regular at $R=4\lambda$, it is easy to show that the EOM (\ref{e1}) is automatically satisfied for any constant curvature background $g_{\mu\nu}$ with $R=4\lambda$. Based on this observation we can easily find solutions to a large variety of $f(R)$ gravities  depending on different forms of $h(R)$. Here we focus on the  simplest case, i.e. choosing $h(R)=1$. The corresponding action is 
\be\label{e2}
\mathcal{S} =\frac{1}{2\kappa^2}\int d^{d+1}x \sqrt{-g} \big[R-2\lambda+\alpha R^2\big]
\ee
where $\alpha\lambda=-1/8$
and the equation of motion is 
\be\label{eom1}
R_{\mu\nu}-\frac{1}{2}g_{\mu\nu}R+\lambda g_{\mu\nu}+2\alpha (g_{\mu\nu}\nabla^2-\nabla_\mu\nabla_\nu)R+2\alpha R(R_{\mu\nu}-\frac{1}{4}g_{\mu\nu}R)= 0.
\ee
Though we have chosen $\alpha\lambda=-1/8$, one can take $\alpha$ and $\lambda$ as independent constants to look for spatially homogeneous Lifshitz black hole solutions, and it turns out that we can have such kind of solutions only at the special coupling with $\alpha\lambda=-1/8$.  

\subsection{Solutions}

Here we will focus on the $d=4$ case. It is known that there are nine classes of solutions by the Bianchi classification. In these nine classes of solutions, the structure constants associated with $\omega^i$ are different and we have different expressions for $\omega^i$ in a particular coordinate basis. For a detailed discussion on these nine Bianchi classes please refer to \cite{Iizuka:2012iv,{Ryan:1975jw}}.  The solutions we are going to look for are asymptotic to the form (\ref{geo}) with $\eta_{ij}$ being diagonal, i.e. $\eta_{ij}=(\lambda_1,\lambda_2,\lambda_3).$ Furthermore, we assume that there is a scaling symmetry associated with the $x_i$ coordinates. 
For simplicity, we will assume $\lambda_{i}$ to be unity in most cases. We closely follow the convention of \cite{Iizuka:2012iv,{Ryan:1975jw}} and  in the following we will give the solutions of the nine cases respectively in $R^2$ gravity. 

\vskip 0.2cm

{$\bullet$ \bf Type I}

This is the simplest case with $c^{i}_{~jk}=0$ and we can set $\omega^i=dx_i~ (i=1,2,3).$  One can choose the following ansatz 
\bea
ds^2&=&L^2\bigg(-e^{2\beta_t r}g(r)dt^2+\frac{dr^2}{g(r)}+e^{2\beta_1 r}dx_1^2+e^{2\beta_2 r}dx_2^2+e^{2\beta_3 r}dx_3^2\bigg),\nonumber\\
g(r)&=&1-M_{-}e^{-\alpha_- r}+M_{+} e^{-\alpha_+ r}. 
\eea
This is the asymptotic Lifshitz solution \cite{Kachru:2008yh,{Adams:2008zk},{Cai:2009ac},{AyonBeato:2010tm}}. Note that we are more interested in the non AdS$_5$ case. From the EOM (\ref{eom1}), we have 
\bea\label{typeIeom}
\alpha_{\pm}&=&\frac{3\beta_t +2(\beta_1+\beta_2+\beta_3)\pm \sqrt{\beta_t ^2+4\beta_t(\beta_1+\beta_2+\beta_3) -4(\beta_1^2+\beta_2^2+\beta_3^2)}}{2},\nonumber\\
\lambda&=&-\frac{1}{8\alpha}=-\frac{1}{2L^2}\big(\beta_t ^2+\beta_t(\beta_1+\beta_2+\beta_3)  +(\beta_1^2+\beta_2^2+\beta_3^2+\beta_1\beta_2+\beta_2\beta_3+\beta_3\beta_1)\big),\nonumber\\
\eea
and  $M_{\pm}$ are arbitrary constants.  We have a large variety of solutions where $\alpha_{\pm}, \beta_t, \beta_i$ obey the relations in (\ref{typeIeom}). 
We can see that this is a black hole solution with two horizons and the boundary is at $r\to \infty$. It can be checked that at $r\to -\infty$ there is a curvature singularity. The thermodynamic properties will be discussed in the next section. 

When $r\to \infty$, it is asymptotic to Lifshitz solution with Lifshitz scaling symmetry 
\be
r\to r+\epsilon,~~~
t\to te^{-\beta_t \epsilon},~~~x_i\to x_i e^{-\beta_i\epsilon},~~~(i=1,2,3).
\ee

\vskip 0.2cm
{$\bullet$ \bf Type II}

In this case the only non-trivial elements of the structure constants are $c^{1}_{~23}=-c^{1}_{~32}=1$.
The invariant one forms are
\be
\omega^1=dx_2-x_1 dx_3,~~~ \omega^2=dx_3,~~~ \omega^3=dx_1,
\ee
The ansatz for the metric is 
\bea
ds^2&=&L^2\bigg(-e^{2\beta_t r}g(r)dt^2+\frac{dr^2}{g(r)}+e^{2(\beta_2+\beta_3)r}(\omega^1)^2+
e^{2\beta_2 r}(\omega^2)^2+e^{2\beta_3 r}(\omega^3)^2
\bigg),\nonumber\\
g(r)&=&1-M_{-}e^{-\alpha_- r}+M_{+} e^{-\alpha_+ r}.
\eea
Solving the Einstein equation (\ref{eom1}) gives 
\bea
\alpha_{\pm}&=&\frac{3\beta_t +4(\beta_2+\beta_3)\pm \sqrt{\beta_t ^2+8\beta_t\beta_2-8\beta_2^2+8\beta_t\beta_3-8\beta_2\beta_3-8\beta_3^2}}{2},\nonumber\\
\lambda&=&-\frac{1}{8\alpha}=-\frac{1}{8L^2}\big[1+4(\beta_t^2+3\beta_2^2 +5\beta_2\beta_3+3\beta_3^2+2\beta_1(\beta_2+\beta_3))\big].
\eea

When $r\to \infty$, it is asymptotic to the solution with the scaling symmetry 
\be
r\to r+\epsilon,~~
t\to te^{-\beta_t \epsilon},~~x_1\to x_1 e^{-\beta_3\epsilon},
~~x_2\to x_2 e^{-(\beta_2+\beta_3)\epsilon},
~~x_3\to x_3 e^{-\beta_2\epsilon}.
\ee 

\vskip 0.2cm
{$\bullet$ \bf Type VI, III and V}

In these three cases,  the non-trivial elements of the structure constants are 
$c^{1}_{~13}=-c^{1}_{~31}=1$ and $c^{2}_{~23}=-c^{2}_{~32}=h$. 
The invariant one forms are
\be
\omega^1=e^{-x_1} dx_2,~~~ \omega^2=e^{-hx_1} dx_3,~~~ \omega^3=dx_1.
\ee
$h=0$ is the type III class and $h=1$ is the type V class. In the type VI class, $h$ can be any constant other than $0,1$. 
As pointed out in \cite{Iizuka:2012iv}, one can take Type III and V as a limit of Type VI. 
We consider the metric ansatz for these three cases 
\bea
ds^2&=&L^2\bigg(-e^{2\beta_t r}g(r)dt^2+\frac{dr^2}{g(r)}+e^{2\beta_1r}(\omega^1)^2+
e^{2\beta_2 r}(\omega^2)^2+(\omega^3)^2
\bigg),\nonumber\\
g(r)&=&1-M_{-}e^{-\alpha_- r}+M_{+} e^{-\alpha_+ r}. 
\eea

For Type VI, we have 
\bea\label{solvi}
\alpha_{\pm}&=&\frac{3\beta_t +2(\beta_1+\beta_2)\pm \sqrt{\beta_t ^2+4\beta_t(\beta_1+\beta_2)-4\beta_1^2-4\beta_2^2}}{2},\nonumber\\
\lambda&=&-\frac{1}{8\alpha}=-\frac{1}{2L^2}\big[\beta_t^2+\beta_t(\beta_1+\beta_2)+\beta_1^2 +\beta_2^2+\beta_1 \beta_2+1+h+h^2\big].
\eea
When $r\to \infty$, it is asymptotic to the solution with  scaling symmetry 
\be
r\to r+\epsilon,~~
t\to te^{-\beta_t \epsilon},~~x_1\to x_1 ,
~~x_2\to x_2 e^{-\beta_1\epsilon},
~~x_3\to x_3 e^{-\beta_2\epsilon}.
\ee 

The solution for Type V is the limit $h\to 1$ of (\ref{solvi}), and Type III is the limit $h\to 0$.

\vskip 0.2cm
{$\bullet$ \bf Type IV}

In this case, $c^{1}_{~13}=-c^{1}_{~31}=1$, $c^{1}_{~23}=-c^{1}_{~32}=1$, $c^{2}_{~23}=-c^{2}_{~32}=1$ and the rest are zero.  The invariant one forms are 
\be
\omega^1=e^{-x_1}dx_2-x_1e^{-x_1}dx_3,~~
\omega^2=e^{-x_1} dx_3,~~
\omega^3=dx_1.
\ee
The metric ansatz is 
\bea
ds^2&=&L^2\bigg(-e^{2\beta_t r}g(r)dt^2+\frac{dr^2}{g(r)}+e^{2\beta_1 r}\big((\omega^1)^2+
(\omega^2)^2\big)+ (\omega^3)^2
\bigg),\nonumber\\
g(r)&=&1-M_{-}e^{-\alpha_- r}+M_{+} e^{-\alpha_+ r}.\eea
We have the solution 
\bea
\alpha_{\pm}&=&\frac{3\beta_t+4\beta_1\pm\sqrt{\beta_t^2+8\beta_t\beta_1-8\beta_1^2}}{2},\nonumber\\
\lambda=-\frac{1}{8\alpha}&=&-\frac{1}{8L^2}\big[13+4\beta_t^2+8\beta_t\beta_1+12\beta_1^2\big].
\eea
When $r\to \infty$, it is asymptotic to  the solution with  scaling symmetry  
\be
r\to r+\epsilon,~~
t\to te^{-\beta_t \epsilon},~~x_1\to x_1,~~x_2\to x_2 e^{-\beta_1\epsilon},~~x_3\to x_3 e^{-\beta_1\epsilon}.
\ee

\vskip 0.2cm
{$\bullet$ \bf Type VII}

In this case, $c^{1}_{~23}=-c^{1}_{~32}=-1$, $c^{2}_{~13}=-c^{2}_{~31}=1$, $c^{2}_{~23}=-c^{2}_{~32}=h$ with $h^2<4$ and the rest are zero.  The special $h=0$ case was considered in \cite{Iizuka:2012iv} and here we focus on the general case. 
 Following \cite{Ryan:1975jw},  we introduce $k=h/2$, $a=\sqrt{1-k^2}$ and  
\be
A=e^{kx_1}\cos{a x_1}, ~~ B=-\frac{1}{a}e^{kx_1}\sin{ax_1},~~ C=e^{-k x_1}\cos{a x_1},~~ D=-\frac{1}{a} e^{-k x_1} \sin{a x_1}.
\ee
The invariant one forms are 
\be
\omega^1=(C-k D) dx_2-D dx_3,~~
\omega^2=D dx_2+(C+k D) dx_3,~~
\omega^3=dx_1.
\ee
The ansatz for the metric is 
\bea
ds^2&=&L^2\bigg(-e^{2\beta_t r}g(r)dt^2+\frac{dr^2}{g(r)}+e^{2\beta_1 r}\big(
(\omega^1)^2+\lambda_1^2 (\omega^2)^2\big)+(\omega^3)^2
\bigg),\nonumber\\
g(r)&=&1-M_{-}e^{-\alpha_- r}+M_{+} e^{-\alpha_+ r}.\eea
The solution is given as 
\bea
\alpha_{\pm}&=&\frac{3\beta_t+4\beta_1\pm\sqrt{\beta_t^2+8\beta_t\beta_1-8\beta_1^2}}{2},\nonumber\\
\lambda=-\frac{1}{8\alpha}&=&-\frac{1}{8\lambda_1^2L^2}\big[1+2(-1+2h^2+2\beta_t^2+4\beta_t\beta_1+6\beta_1^2)\lambda_1^2+\lambda_1^4\big].
\eea
When $r\to \infty$, it is asymptotic to  the solution with  scaling symmetry 
\be
r\to r+\epsilon,~~
t\to te^{-\beta_t \epsilon},~~x_1\to x_1 ,~~x_2\to x_2 e^{-\beta_1\epsilon},~~x_3\to x_3 e^{-\beta_1\epsilon}.
\ee

\vskip 0.2cm
{$\bullet$ \bf Type IX and Type VIII}

In the type IX class, $c^{1}_{~23}=-c^{1}_{~32}=1$, $c^{2}_{~31}=-c^{2}_{~13}=1$, $c^{3}_{~12}=-c^{3}_{~21}=1$ and the others are zero. The invariant one forms are: 
\bea\label{ixomega}
\omega^1&=&-\sin x_3 dx_1+\sin x_1 \cos x_3 dx_2,\nonumber\\ 
\omega^2&=&\cos x_3 dx_1+\sin x_1\sin x_3 dx_2,\nonumber\\
\omega^3&=&\cos x_1dx_2+dx_3.
\eea
We choose the metric to be 
\bea
ds^2&=&L^2\bigg(-e^{2\beta_t r}g(r)dt^2+\frac{dr^2}{g(r)}+(\omega^1)^2+
(\omega^2)^2+\lambda_1 (\omega^3)^2
\bigg),\nonumber\\
g(r)&=&1-M_{-}e^{-\alpha_- r}+M_{+} e^{-\alpha_+ r}.
\eea
The solution is
\bea
\alpha_{+}&=&2\beta_t,~~~\alpha_{-}=\beta_t \nonumber\\
\lambda&=&-\frac{1}{8\alpha}=-\frac{1}{8L^2}\big[4\beta_t^2-4+\lambda_1\big].
\eea
When $r\to \infty$, it is asymptotic to  the solution with  scaling symmetry 
\be
r\to r+\epsilon,~~
t\to te^{-\beta_t \epsilon},~~x_i\to x_i,~(i=1,2,3).
\ee 
This geometry is asymptotically AdS$_2$ $\times$ squashed $S^3$ \cite{Iizuka:2012iv} and it can be viewed as  AdS$_2$ black  hole $\times$ squashed $S^3$. It is well known \cite{{Spradlin:1999bn},Faulkner:2011tm} that the AdS$_2$  black hole is locally  equivalent to the vacuum AdS$_2$. However, we can still define horizons and it is interesting to study it. Similar things also happen to the next case. 

Bianchi Type VIII case is quite similar to Type IX case. 
The type VIII case is $c^{1}_{~23}=-c^{1}_{~32}=-1$, $c^{2}_{~31}=-c^{2}_{~13}=1$ and $c^{3}_{~12}=-c^{3}_{~21}=1$ and the others are zero. The invariant one form are 
\bea
\omega^1&=&dx_1+(1+x_1^2)dx_2+(x_1-x_2-x_1^2x_2)dx_3,\nonumber\\ 
\omega^2&=&2x_1 dx_2+(1-2x_1 x_2) dx_3,\nonumber\\
\omega^3&=&dx_1+(-1+x_1^2)dx_2+(x_1+x_2-x_1^2x_2)dx_3.
\eea
Similar to Type IX case,  the ansatz of the metric  is
\bea
ds^2&=&L^2\bigg(-e^{2\beta_t r}g(r)dt^2+\frac{dr^2}{g(r)}+(\omega^1)^2+
(\omega^2)^2+\lambda_1 (\omega^3)^2
\bigg),\nonumber\\
g(r)&=&1-M_{-}e^{-\alpha_- r}+M_{+} e^{-\alpha_+ r}.
\eea
The solution is
\bea
\alpha_{+}&=&2\beta_t,~~~\alpha_{-}=\beta_t \nonumber\\
\lambda&=&-\frac{1}{8\alpha}=-\frac{1}{8\lambda_1L^2}\big[4\beta_t^2\lambda_1+4+\lambda_1^2\big].
\eea

\vskip 0.2cm

After exhibiting these solutions for the nine cases, we can make several interesting comments.  
First, though $\alpha/L^2$ is still small compared to $1/\lambda L^2$, these solutions only exist at a specific value of $\alpha$ and we can not take the solutions as a perturbation around pure Einstein solutions according to $\alpha$.
Second, here we  have the constraint that $\alpha_{\pm}$ are real which would give more restrictions on the parameters appearing in the solution.  Third,  in some of them such as type II, for the special couplings with $\alpha_+=\alpha_-$, the two sectors in $g(r)$ are the same and a new logarithmic solution will appear, just like what happens in three dimensional massive gravity case \cite{Grumiller:2008qz,{Liu:2009pha}}. Actually, one can choose $M_{\pm}$ with a special relation and take the $\alpha_+\to \alpha_-$ to construct these logarithmic solutions like \cite{AyonBeato:2010tm}. 

\subsection{Thermodynamic properties}

One can choose suitable $M_{\pm}$ and parameters like $\beta_t$ to make the solution we obtained a black hole solution. 
Following the discussions in \cite{Cai:2009ac}  the thermodynamic properties of these black hole solutions can be discussed in parallel. 

The horizon of these black holes is located at $r_H$ where $g(r_H)=0.$ In general, there might be two horizons and we will consider the 
non-degenerate case\footnote{The degenerate case corresponds to vacuum solution or the extremal black holes, e.g. in the logarithmic black hole case \cite{AyonBeato:2010tm}. Surely it is interesting to study them to see whether they are more stable than the vacuum solutions.} and choose $r_H$ as the location of the outer horizon. We have $g(r)\sim g'(r_H)(r-r_H)$ when $r\to r_H$.   It is easy to show that the Hawking temperature of the black hole is 
\be\label{tem}
T =\frac{g'(r_H)}{4\pi }e^{\beta_t r_H}.
\ee
From Wald's formula for the black hole entropy~\cite{Wald},
\begin{equation}
S =- 2\pi\int_{\text{H}}d\Sigma_H \frac{\partial {\cal L}}{\partial
R_{\mu\nu\rho\sigma}}\epsilon_{\mu\nu}\epsilon_{\rho\sigma} =-\frac{\pi}{2\kappa^2} \int_{\text{H}}d\Sigma_H (1+2\alpha R)(g^{\mu\rho}g^{\nu\sigma}-g^{\mu\sigma}g^{\nu\rho})\epsilon_{\mu\nu}\epsilon_{\rho\sigma} ,
\end{equation}
where $\epsilon_{\mu\nu}$ is the binormal to the horizon and is normalized as $\epsilon^{\mu\nu}\epsilon_{\mu\nu} = -2$. We find that all the above constant curvature black hole solutions  have a vanishing
entropy in the action (\ref{e2}). 

 Furthermore, the on shell action  (\ref{e2}) is always vanishing for these constant curvature solutions  
 with $R=4 \lambda=-1/2\alpha $, even when a boundary term \cite{Cremonini:2009ih} is added
\begin{equation}
{\mathcal S}_{\rm bt}=-\frac{1}{8\pi} \int_{\partial \cal M}d^4x
\sqrt{-h}(1+2\alpha R)K,
\end{equation}
where $K$ is the extrinsic curvature for the boundary hypersurface
${\partial \cal M}$ with induced metric $h$.  This implies that both
the free energy and mass of the black holes vanish as well
\begin{equation}
F=M=0,
\end{equation}
although the black hole solutions have a nonvanishing horizon radius $r_H$.

Such kinds of zero entropy and zero mass black holes have been found before in critical NMG for BTZ black holes where the corresponding field theory sensitively depends on the boundary condition \cite{{Liu:2009bk},{Liu:2009kc}}.  (See also  \cite{{Cai:2009ac}, Cai:2009de,{Lu:2011zk},{Deser:2011xc}}.)  It is natural to expect  that the dual field theory (if exists) to each fixed asymptotical geometry provides a simple consistent toy model for quantum gravity in five dimensions.

\section{\label{s3} General quadratic curvature gravity}

In this section, we will consider the spatially homogeneous Lifshitz black holes in the most general  quadratic curvature gravity. The action is 
\be\label{s3e1}
\mathcal{S} =\frac{1}{2\kappa^2}\int d^{5}x \sqrt{-g} \bigg[R-2\lambda+\alpha R^2+\beta R_{\mu\nu} R^{\mu\nu}+\gamma
\big(R_{\mu\nu\rho\sigma}R^{\mu\nu\rho\sigma}-4R_{\mu\nu}R^{\mu\nu}+R^2\big) \bigg] \ee
where $\alpha, \beta, \gamma$ are coupling constants and $\lambda$ is the cosmological constant.  Note that the term associated to $\gamma$ is the Gauss-Bonnet term.  The corresponding equation of motion for the action (\ref{s3e1}) is 
\bea\label{s3e2}
&&R_{\mu\nu}-\frac{1}{2}g_{\mu\nu}R+\lambda g_{\mu\nu}+(2\alpha+\beta)(g_{\mu\nu}\nabla^2-\nabla_\mu\nabla_\nu)R+\beta(\nabla^2 R_{\mu\nu}-\frac{1}{2}g_{\mu\nu}\nabla^2 R)
\nonumber\\
&&  +2\alpha R(R_{\mu\nu}-\frac{1}{4}g_{\mu\nu}R)
+2\beta(R_{\mu\rho\nu\sigma}-\frac{1}{4}g_{\mu\nu}R_{\rho\sigma})R^{\rho\sigma}+2\gamma\big(RR_{\mu\nu}-2R_{\mu\rho\nu\sigma}R^{\rho\sigma}
\nonumber\\
&&
+R_{\mu\rho\sigma\tau}R_{\nu}^{~\rho\sigma\tau}-2R_{\mu\rho}R_{\nu}^{~\rho}
-\frac{1}{4}g_{\mu\nu}(R_{\rho\sigma\tau\ell}R^{\rho\sigma\tau\ell}-4R_{\rho\sigma}R^{\rho\sigma}+R^2)\big)=0
\eea

The action considered in the previous section is a special case for this action with $\alpha=-1/8\lambda,\beta=0, \gamma=0$.
One can rescale $g_{\mu\nu}\to L^2g_{\mu\nu}$, correspondingly, $\lambda\to L^{-2} \lambda, (\alpha,\beta,\gamma)\to L^2(\alpha,\beta,\gamma)$ to set $L=1$. It is easy to show that the solutions for the cases of 
Type II, VI, III, V, IV, VII and VIII are the same as in the previous section, i.e. in these cases we can only find solutions at 
$\beta=\gamma=0$ and $\alpha \lambda=-1/8$.   For other cases, we can have new solutions where either $\beta\neq 0$ or $\gamma\neq 0$. We list our results in the following: 

\vskip 0.2cm

{$\bullet$ \bf Type I}

In this case we focus on the vacuum solutions and the metric ansatz is the following
\be
ds^2=-e^{2\beta_t r}  dt^2+dr^2+e^{2\beta_1 r}dx_1^2+e^{2\beta_2 r}dx_2^2+e^{2\beta_3 r}dx_3^2.\ee

For $\beta_1=\beta_2=\beta_3$, we have the solution 
\be
\lambda=-\frac{1}{2}\big(\beta_t ^2+3\beta_t\beta_1 +6\beta_1^2-4\gamma\beta_t\beta_1^2(\beta_t+3\beta_1)\big),~~
\alpha=-\frac{2\beta(3\beta_1^2+\beta_t^2)+4\gamma\beta_1^2-1}{4(\beta_t^2+3\beta_t\beta_1+6\beta_1^2)}.
\ee
Note that here we are more interested in the non AdS$_5$ case.
For $\alpha=\beta=0$, the theories reduce to 
Einstein Gauss-Bonnet gravity. We can see that there are Lifshitz solutions  for Einstein Gauss-Bonnet gravity besides the asymptotically  AdS solutions found in  \cite{Cai:2001dz}. 
Here we focus on the vacuum solutions. The corresponding black hole solutions 
have been studied in \cite{AyonBeato:2010tm} and their thermodynamics can be 
found in \cite{Devecioglu:2011yi}.  Also note that with and without the accompanying Proca matter field in this case the asymptoticallly Listshitz black hole solutions were obtained and their corresponding thermodynamic properties were studied in \cite{Dehghani:2010kd, {Dehghani:2010gn},{Brenna:2011gp}}. 

For general $\beta_i$ $(i=1,2,3)$
we have 
\bea
\lambda&=&-\frac{1}{2}\big(\beta_t^2+\beta_t(\beta_1+\beta_2+\beta_3)+\beta_1^2+\beta_2^2+\beta_3^2+\beta_1\beta_2+\beta_2\beta_3+\beta_3\beta_1\big),\nonumber\\
\alpha&=&\frac{1-2\beta (\beta_t^2+\beta_t(\beta_1+\beta_2+\beta_3)+\beta_1^2+\beta_2^2+\beta_3^2+\beta_1\beta_2+\beta_2\beta_3+\beta_3\beta_1)}{4(\beta_t^2+\beta_1^2+\beta_2^2
+\beta_3^2)},\nonumber\\
\gamma&=&0.
\eea

From the above formula, we can see that if we take $\beta=\gamma=0$, we will come back to the results in the last section. 

\vskip 0.2cm

{$\bullet$ \bf Type IX}

In this case the ansatz for the black hole metric is 
\bea
ds^2&=&-e^{2\beta_t r}g(r)dt^2+\frac{dr^2}{g(r)}+(\omega^1)^2+
(\omega^2)^2+\lambda_1 (\omega^3)^2,\nonumber\\
g(r)&=&1-M_{-}e^{-\alpha_- r}+M_{+} e^{-\alpha_+ r} 
\eea
where $\omega^i$ is defined in (\ref{ixomega}). The EOM (\ref{s3e2}) gives: 
\bea
\lambda&=&\frac{-8\beta_t^6+2(5+4\gamma)\beta_t^4\lambda_1+\beta_t^2[8+\lambda_1(3\lambda_1-2\gamma(\lambda_1-4)-10)]+\lambda_1(\lambda_1-4)}{8\beta_t^2(2\beta_t^2-3\lambda_1+2)-8\lambda_1},\nonumber\\
\alpha&=&\frac{-8\gamma\beta_t^6+2\beta_t^4+\beta_t^2[2-3\lambda_1+\gamma(\lambda_1-2)(3\lambda_1-4)]-\lambda_1}{[2\beta_t^4+\beta_t^2(2-3\lambda_1)-\lambda_1](4\beta_t^2+\lambda_1-4)},\nonumber\\
\beta&=&\frac{\gamma\beta_t^2(4\beta_t^2+4-\lambda_1)}{2\beta_t^4+\beta_t^2(2-3\lambda_1)-\lambda_1},
\nonumber\\
\alpha_{+}&=&2\beta_t,~~~\alpha_{-}=\beta_t.\eea
Similar to the previous case, we can choose $\beta_t=4+\frac{1}{\gamma},\lambda_1=\frac{1}{2\sqrt{\gamma}}$ to make $\alpha=\beta=0$, i.e. the solution also appears in Einstein Gauss-Bonnet gravity. 
In this case, one can also study the thermodynamic properties. The temperature is the same as (\ref{tem}) in the previous section. By Wald's formula \cite{Wald}, we have 
\bea
S &=&- 2\pi\int_{\text{H}}d\Sigma_H \frac{\partial {\cal L}}{\partial
R_{\mu\nu\rho\sigma}}\epsilon_{\mu\nu}\epsilon_{\rho\sigma} \nonumber\\
&=&-\frac{\pi}{2\kappa^2} \int_{\text{H}}d\Sigma_H\epsilon_{\mu\nu}\epsilon_{\rho\sigma}\bigg( (1+2(\alpha+\gamma) R)(g^{\mu\rho}g^{\nu\sigma}-g^{\mu\sigma}g^{\nu\rho})\nonumber\\
&&~~+(\beta-4\gamma) (g^{\mu\rho}R^{\nu\sigma}+g^{\nu\sigma}R^{\mu\rho}-g^{\mu\sigma}R^{\nu\rho}-g^{\nu\rho}R^{\mu\sigma})+4\gamma R^{\mu\nu\rho\sigma}\bigg)\nonumber\\
&=&-\frac{2\pi\beta\lambda_1\Omega_3}{\kappa^2\beta_t^2},
\eea
where $\Omega_3$ is the volume of the three dimensional space. It is interesting to see that we have a constant entropy  which does not depend on the temperature.  As pointed out in the previous section, this geometry is an  $AdS_2$ black hole $\times$ squashed $S^3$. Essentially the temperature independence of the entropy is a feature of  $AdS_2$ black hole \cite{Spradlin:1999bn} because the temperature can be eliminated by a coordinate transformation.  This might not be true in the case of the quantum theory and it would be interesting to consider the  entanglement entropy etc to study the quantum gravity theory in this geometry background.

\section{Conclusion and Discussion}

As shown in  \cite{Iizuka:2012iv}, we can relax the usual translational invariance to the requirement of  homogeneity for the spatial directions of the black holes. There are nine Bianchi classes of  spatially homogeneous Lifshitz vacuum solutions.  In this paper, we studied the higher derivative gravity theories to get these kinds of  spatially homogeneous Lifshitz black hole solutions.  We  constructed the analytic black hole solution  which asymptotes to the  spatially homogeneous Lifshitz vacuum solutions in both $R^2$ gravity and the most general quadratic curvature gravity theories. We analyzed the thermodynamics of these  spatially homogenous Lifshitz black holes in $R^2$ gravity and found that they have zero entropy at non-zero temperatures which is quite similar to some previous studies \cite{{Liu:2009bk},{Cai:2009ac}, Cai:2009de,{Lu:2011zk}}. New solutions are found for Bianchi Type I and IX models in the more general quadratic curvature gravities.   The type IX solution is essentially  $AdS_2$ black hole $\times$ squashed $S^3$ and its entropy is found to be independent of the temperature as expected.

There are some remaining open questions. One immediate question is to 
study the conserved charges of these spatially homogeneous Lifshitz black holes in higher-curvature gravity 
following e.g. \cite{Hohm:2010jc,{Devecioglu:2010sf}}.  As for applications to condensed matter systems, it is necessary to 
study the scalar or spinor two point functions or entanglement entropy or probe D-branes in these spacetimes.  It would be interesting to construct the full solutions flowing from these spatially homogeneous solutions at the horizon to AdS$_5$ on the boundary and study the corresponding fluid property \cite{Cremonini:2011ej} and this may also give more restrictions to the parameters in the solutions.

While here we considered the pure quadratic curvature gravity, 
one could try to find the spatially homogeneous Lifshitz black holes in other kinds of gravity theories, such as in the Einstein-Maxwell-dilaton gravities \cite{Charmousis:2010zz} or Lovelock gravities etc.  For the Einstein gravity coupled to massive gauge field with higher 
curvature corrections, the spatially homogeneous and isotropic Lifshitz black holes and their theomodynamics were studied in \cite{Dehghani:2010kd,{Dehghani:2010gn},{Brenna:2011gp}}.
Similarly, one can also add the higher derivative gravity corrections (or Chern-Simons corrections) to  \cite{Iizuka:2012iv} to look for spatially homogeneous Lifshitz solutions and study their thermodynamics.  Finally, it would be interesting to study the instability as well as the causal structures of these geometries.

\section*{Acknowledgments}
We would like to thank Rong-Gen Cai, Koenraad Schalm, Ya-Wen Sun and Jan Zaanen for 
interesting discussions and collaboration on related issues.  
This work is supported by the Dutch Foundation for Fundamental Research on Matter (FOM).

\end{document}